\documentclass[conference,letterpaper]{IEEEtran}
\IEEEoverridecommandlockouts
\usepackage{amsmath,bm}
\usepackage[final]{graphicx}
\usepackage{subfigure}
\usepackage{amsthm}
\usepackage{verbatim}
\usepackage{mathrsfs}
\usepackage{euscript}
\usepackage{flushend}
\usepackage{amsfonts}
\usepackage{verbatim}

\newtheorem{theorem}{Theorem}
\newtheorem{lemma}{Lemma}
\newtheorem{corollary}{Corollary}
\newtheorem{proposition}{Proposition}

\theoremstyle{definition}

\newcommand{\rb}{\mathbf r}

\newcommand{\xb}{\mathbf x}

\newcommand{\yb}{\mathbf y}

\newcommand{\Yb}{\mathbf Y}

\newcommand{\defeq}{\overset{\text{\tiny{def}}}{=}}

\begin{document}
%
\title{On the Capacity of the One-Bit \\ Deletion and Duplication Channel}

\author
{\IEEEauthorblockN{Hamed Mirghasemi and Aslan
Tchamkerten}

\thanks{This work was supported in part by an Excellence Chair Grant
from the French National Research Agency (ACE
project). H. Mirghasemi and A. Tchamkerten are
with the Communications
and Electronics Department,
Telecom ParisTech, 75634 Paris Cedex 13. Email: \{mirghasemi,aslan.tchamkerten\}@telecom-paristech.fr.}
}

\maketitle

\begin{abstract}

The one-bit deletion and duplication channel   
is investigated. An input to this channel consists of a block of 
 $\ell\geq 1$ bits  which experiences a deletion with probability $p$,  a duplication with probability $q$, and  remains
unchanged with probability $1-p-q$. For this channel a capacity expression is obtained in the asymptotic regime where $p+q=o(1/\log \ell)$.  As a corollary, we obtain an asymptotic expression for the capacity of the so called ``segmented'' deletion and duplication channel where the input now consists of several blocks and each block independently experiences either a deletion, or a duplication, or remains unchanged.

\end{abstract}

%

\section{Introduction}
Given an integer $\ell\geq 1$ and two constants $p, q \in [0,1]$ such that $p+q \leq 1$, the segmented deletion and duplication 
channel treats independently each consecutive
length $\ell$ binary input block in one of the following ways:
\begin{itemize}
\item  one bit is deleted with probability $p$,
\item one bit is duplicated with probability $q$,
\item the block remains unchanged with probability $1-p-q$.
\end{itemize}
Conditioned on a bit being deleted (duplicated) in a
particular block, the deletion (duplication) occurs randomly and uniformly
over the block. Hence, the unconditional probability that
any particular bit is deleted or duplicated is equal to ${p}/{\ell}$ 
and ${q}/{\ell}$,
respectively.

When $\ell=1$, the segmented deletion and duplication
 channel becomes the standard deletion and duplication
channel where each input bit is 
independently deleted with probability $p$, duplicated with probability $q$, and is left unchanged with probability of $1-q-p$.\footnote{See, {\it{e.g.}},
\cite{Diggavi06,DrineaImp07,Kanoria10,KirschDir10,Rahmati11,Venkataamanan11} 
for recent references on the i.i.d. deletion
and duplication channel.}

An input to the channel consists of $s\geq 1$ consecutive
blocks of length $\ell$. The corresponding output is thus
a binary string of known
length between $n-s$ and $n+s$ where $$n\defeq
s\cdot \ell\,.$$

Rate $R$ is said to be achievable
if, for any $\varepsilon>0$ and $s$ large
enough,
there exist $ 2^{nR}$ codewords and a decoder
whose average error probability over codewords is
no larger than $\varepsilon$. Capacity is the
supremum of achievable rates and admits the asymptotic expression \begin{align}
C=\lim_{s \rightarrow \infty}\frac{1}{n}\max_{X^n}
I(X^n ;\Yb(X^n))\, \label{eq:Dobrushin}
\end{align}
according to Dobrushin's capacity theorem \cite[Theorem 1]{Dobrushin67}.

Segmented channels with synchronization errors were introduced by Liu and Mitzenmacher in \cite{LiuMitzenmacher10} where,
following an algorithmic approach, they proposed 
a zero-error coding
scheme and thereby  established a numerical lower
bound on the capacity of the segmented deletion
channel ({\it{i.e.}}, for $q=0$).

A difficulty in obtaining a tight single-letter
characterization of $C$  stems from the fact that the
receiver does not know the error
pattern, {\it{i.e.}}, which out of the $s$ blocks
experienced a deletion or a duplication (albeit it knows the overall
number of deletions and duplications). As a
consequence, errors ``propagate'' across blocks. 

A useful
technique to derive upper and lower bounds on $C$
is to reveal the receiver the error pattern
$E^s=\{E_i\}_{i=1}^s$ where $E_i=-1$ if the $i$-th
block experienced a deletion, $E_i=1$ if the $i$-th
block experienced a duplication, and $E_i=0$
otherwise \cite{Fertonani10, Wang}. When this side
information is provided to the receiver, each block
can be considered in complete isolation and we obtain
the so-called ``one-bit'' deletion and duplication
channel. The capacity $C_{SI}$ of the one-bit
deletion and duplication channel is the capacity with
respect to a single length $\ell$ block. We hence
have the obvious upper bound 
\begin{align} \label{eq:generalUB} C \leq C_{SI}\,,
\end{align} where \begin{align}\label{eq:dobru}
C_{SI}=\frac{1}{\ell}\max_{X^{\ell}}{{I}(X^{\ell};{\Yb}(X^{\ell}))}\,,
\end{align}  where $X^\ell$ denotes a random input
block to the channel, and where 
${\Yb}(X^{\ell})$ denotes  the corresponding
output.

A lower bound to $C$ in terms of $C_{SI}$ can be
obtained by using the 
argument of \cite[Section II.C]{Wang}. First observe
that $$I(X^n ;\Yb(X^n),E^s)\leq I(X^n
;\Yb(X^n))+H(E^s)\,.$$
Using that $H(E^s)=s H_b(p,q)$  where $H_b(p,q)$ denotes\footnote{Logarithms
are taken to the base $2$ throughout the paper.} the entropy function
$-p\log{p}-q\log{q}-(1-p-q)\log{(1-p-q)}\,,$
 it then follows that
\begin{align}\label{eq:lboundctilde}
 C_{SI}-\frac{1}{\ell}H_b(p,q)\leq {C}\,.
 \end{align}
Note that an analytical expression for
$C_{SI}$ remains to be found and a
numerical evaluation, for instance, via the Arimoto-Blahut algorithm, is computationally heavy
already for moderate values of $\ell$, say
$\ell\geq 17$. 

 In this paper, we provide analytical upper and lower bounds on
$C_{SI}$ which, via \eqref{eq:generalUB} and
\eqref{eq:lboundctilde}, yield upper and lower
bounds on $C$. These bounds are tight in certain asymptotic regimes yielding the main capacity results.

Throughout the paper, the following notational conventions are
adopted. A binary length $n$ vector is usually denoted
by a bold script, {\it{e.g.}}, $\xb$, and
its length is denoted by $|\xb|$. If we
want to emphasize the length of a vector, we 
alternatively write
$x^n$. 
For computational convenience, we sometimes refer to a particular sequence
$\xb$ using its runlength description
$\rb(\xb)=(x_1,  \{r_i(\xb)\})$ where $r_i(\xb)$ denotes its $i$\-th runlength.\footnote{Notice that $\sum_i
r_i(\xb)=|\xb|$.} For instance, the runlength
description of $0100110$ is $(0,11221)$. 

We use $\yb\prec \xb$ whenever
$\yb$ is a subsequence of $\xb$, {\it{i.e.}}, whenever $\yb$ results from
the deletions of $|\xb|-|\yb|$ bits of $\xb$.

The next 
section contains our main  results and
Section \ref{section3} is devoted to the proofs.

\section{Main Results} \label{section2}

Let
\begin{align}\label{lalphasi}
{L}_{SI}^\alpha \defeq \frac{I(X^{\ell}(\alpha);{\Yb}(X^{\ell}(\alpha)))}{\ell}
\end{align}
where $X^{\ell}(\alpha)=X_1,X_2,\ldots, X_\ell$ refers to the Markovian input given by
\begin{align}\label{eq:DisMarkov}
&Pr(X_1=0)=Pr(X_1=1)=\frac{1}{2} \nonumber \\
&Pr(X_i \neq X_{i-1})=\alpha, \;   \; 2 \leq i \leq \ell\,,
\end{align}
for some fixed parameter $\alpha\in [0,1]$.

An explicit expression for  the lower bound \eqref{lalphasi}  in terms of the parameters $\ell$,
$p$, $q$, and $\alpha$ is given in the appendix.

Further, define
\begin{align}\label{eq:Up_analytical}
U\defeq & \frac{p \cdot(\ell-1)+q \cdot \log{(2^{\ell+1}-2})}{\ell}\nonumber\\
&+\frac{(1-p-q)\log{\sum_{x^{\ell}\in\{0,1\}^{\ell}}{2^{-\frac{p+q}{1-p-q}{\hat{H}({\rb(x^{\ell})})}}}}}{\ell} \,,
\end{align}
where $\hat{H}(\rb(x^{\ell}))$ is the runlength
empirical entropy of $x^{\ell}$ 
\begin{align}
\hat{H}(\rb(x^{\ell}))\defeq
-\sum_{i\geq 1}{\frac{r_i(x^\ell)}{\ell}\log{\frac{r_i(x^\ell)}{\ell}}}\,. \nonumber 
\end{align}

\begin{proposition}\label{Thm:CapacityMarkov}
For any $p, q, \alpha \in [0,1]$ such that $p+q \leq 1$ and any integer $\ell> 1$, we have
\begin{align}\label{eq:MarLowDelDup}
{L}_{SI}^\alpha \leq
C_{SI}\leq {U}\,.
\end{align} 
\end{proposition}

\begin{figure}
    \includegraphics[scale=.26]{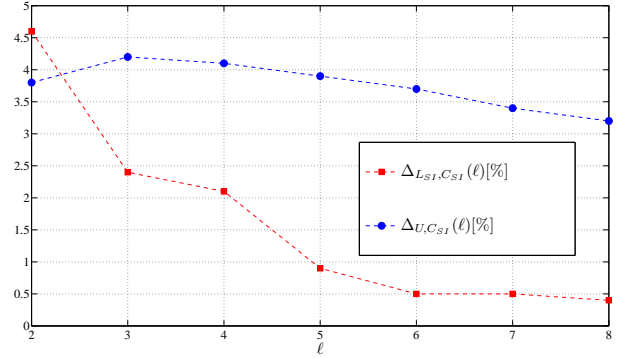}
    \vspace{-0.9cm}
    \caption{Relative differences between $C_{SI}$ and its upper bound $U$ (given by $ \Delta_{U,C_{SI}} (\ell)$) and between $C_{SI}$ and its lower bound $\max_\alpha{L}_{SI}^\alpha$ (given by $\Delta_{{L}_{SI},C_{SI}} (\ell)$).}
    \label{fig:delats_low_up}
\end{figure}

In Fig.~\ref{fig:delats_low_up},
\begin{align}
 \Delta_{U,C_{SI}} (\ell)&\defeq \max_{p,q:p+q \leq 1}
\frac{U-C_{SI}}{C_{SI}}
 \end{align} 
and
 \begin{align}
\Delta_{{L}_{SI},C_{SI}} (\ell)&\defeq \max_{p,q:p+q \leq 1}
\frac{C_{SI}-\max_{\alpha}{{L}_{SI}^\alpha}}{C_{SI}}
\end{align}
represent the relative difference between $C_{SI}$, which is obtained numerically by the Arimoto-Blahut algorithm, and the upper and
 lower bounds $U$ and ${L}_{SI}^\alpha$, respectively,
 the latter being numerically optimized over
 $\alpha\in [0,1]$.

As we can see, these bounds are fairly
close for a wide range of $p$ and $q$. For instance,
their difference with respect to $C_{SI}$ is
at most $5\%$ for any $p$ and $q$ such that $p+q\leq 0.6$, as long as $\ell\geq
2$. Moreover,  numerical evidence suggests that both
$\Delta_{U,C_{SI}} (\ell)$ and $\Delta_{{L}_{SI},C_{SI}} (\ell)$ tend to zero as $\ell\to \infty$.  

\begin{figure}
    \includegraphics[scale=.26]{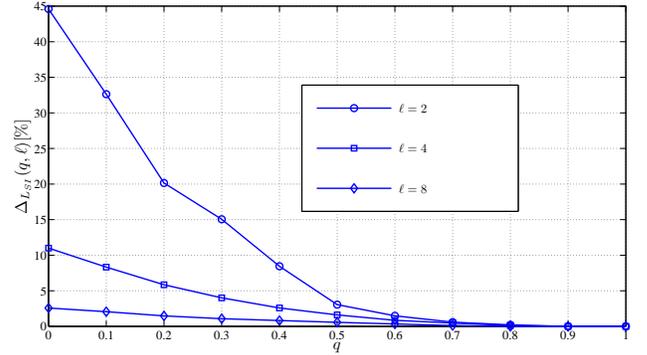}
    \vspace{-0.9cm}
    \caption{Relative difference between $\max_\alpha{L}_{SI}^{\alpha}$ and ${L}_{SI}^{0.5}$. }
    \label{fig:low_rel_dist}
\end{figure}

In Fig.~\ref{fig:low_rel_dist} 
\begin{align}
\Delta_{L_{SI}}(q,\ell) \defeq  \max_{p \in [0,1-q]}\frac{\max_{\alpha \in [0,1]}{{L}_{SI}^{\alpha}}-{L}_{SI}^{0.5}}{\max_{\alpha \in [0,1]}{{L}_{SI}^{\alpha}}}
\end{align}
represents the relative difference between ${L}_{SI}^{0.5}$ and the optimized
lower bound expression 
$\max_{\alpha}{{L}^{\alpha}_{SI}}$ as a function of
$q$, for different
values of $\ell$. As we observe, when either $\ell$ or $q$ decreases, non-uniform inputs perform significantly better than uniform inputs.

We now turn to the case where there is no side information at the receiver.  For comparing our results with related work, we restrict ourselves to the purely deletion case, {\it{i.e.}}, $q=0$. For this channel, a lower bound to capacity is obviously 
$$L^\alpha\defeq L^\alpha_{SI}-H_b(p,q)/\ell$$
by  \eqref{eq:lboundctilde} and \eqref{eq:MarLowDelDup}.  

\begin{figure}
    \includegraphics[scale=.27]{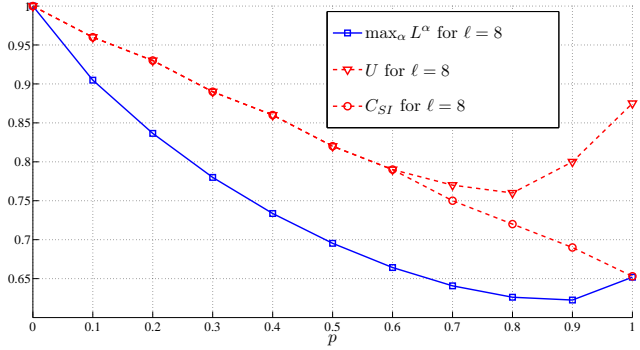}
    \vspace{-0.9cm}    
    \caption{Upper and lower bounds on the capacity of segmented deletion channel for $\ell=8$.}
    \label{fig:FigDall8}
\end{figure}

\begin{figure}
    \includegraphics[scale=.27]{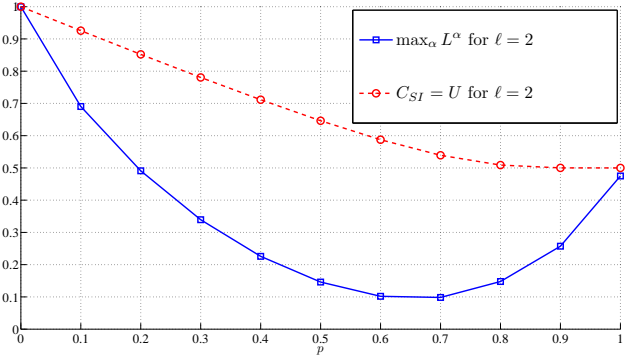}
    \vspace{-0.9cm}
    \caption{Upper and lower bounds on the capacity of segmented deletion channel for $\ell=2$.}
    \label{fig:FigLUtC}
\end{figure}

Figures~\ref{fig:FigDall8} and \ref{fig:FigLUtC}  represent the upper and lower bounds on $C$ given by $U$ and $\max_{\alpha}L^{\alpha}$ for $\ell=8$ and  $\ell=2$, respectively. The difference between these bounds is particularly significant for $p\approx 1/2$. Indeed, this is partly due to the fact that the difference between the two bounds is lower by the side information $H_b(p,0)/\ell$ which is maximal for $p=1/2$.   
Also note that
$U$  may be
better or worse than the numerical upper bound
given in \cite{Wang}. For instance, for  $\ell=8$ (Fig.~\ref{fig:FigDall8}) we have
that  $ U$ is  lower than the upper bound proposed in \cite{Wang} for $ p \in
[0,0.6]$ whereas the opposite
holds for $ p \in
(0.6,1]$. Finally note that $U$ appears to be a very good approximation for $C_{SI}$; the difference gets negligible for $p\leq 0.6$ when $\ell=8$ and is negligible for any $p\leq 1$ when $\ell=2$.
\subsection*{Asymptotics}
In the regime of large blocks and small synchronization errors we have:\footnote{We say that $f(\ell)=O(g(\ell))$ if there exists a positive real number $k$ such that $|f(\ell)| \leq k\cdot g(\ell)$ when $\ell \to \infty$.} 
\begin{theorem} \label{Thm:MarLowBoundAsymp} 

\begin{itemize}
\item[i.] For $p$ and $q$ such that $p+q \leq 1$, we
have
\begin{align}\label{eq:asymtilde}
{L}_{SI}^{0.5}=&1-\frac{p+q}{\ell}\log{\ell}+\frac{p}{\ell}(K-1)+\frac{q}{\ell}(K+1)\nonumber \\
&+(p+q)O(\ell^{-2})\,;
\end{align}
\item[ii.] When $(p+q)\log{\ell} \to 0$, we have
\begin{align}\label{eq:asymup}
U=&1-\frac{p+q}{\ell}\log{\ell}+\frac{p}{\ell}(K-1)+\frac{q}{\ell}(K+1)\nonumber\\
&+O((p+q)^2 (\log{\ell})^2/\ell)\,;
\end{align}
where $K=\sum_{j=1}^{\infty}{\frac{j
\log{j}}{2^{j+1}}} \simeq 1.2885$.\footnote{This
constant appeared as $A_1$ in \cite[Theorem
1]{KanoriaOptimal11}.}
\item[iii.]
When $(p+q)\log{\ell} \to 0$, we have
\begin{align}\label{eq:Asymp_cap}
C_{SI}=&1-\frac{p+q}{\ell}\log{\ell}+\frac{p}{\ell}(K-1)+\frac{q}{\ell}(K+1)\nonumber\\
&+(p+q)O(\ell^{-2})+O(\frac{(p+q)^2}{\ell} \log^2{\ell})\,.
\end{align}
\end{itemize}
\end{theorem}
We note that for $p=1$ (and hence $q=0$),
the $1-\frac{\log{\ell}}{\ell}$ term in \eqref{eq:Asymp_cap} corresponds to the
zero-error capacity of the  one-bit purely deletion
channel (\cite[Theorem 2.5]{Sloane01}). 

Note that $p$ and $q$ do not play symmetric roles in the asymptotic capacity expression \eqref{eq:Asymp_cap}. An intuitive explanation for this is as follows.  From the length of the output block the decoder knows whether the input to the channel experiences a deletion, a duplication, or remains unchanged. If a duplication occurs,  then the decoder also knows the number of runs in the input since duplication cannot change the number of runs. By contrast, deletion errors can erase a run completely, thereby increasing decoding ambiguity. 
From Theorem~\ref{Thm:MarLowBoundAsymp} and
\eqref{eq:lboundctilde}, we readily obtain the following
  asymptotic expressions for the segmented deletion and duplication channel:
\begin{corollary}\label{cor:asymseg}
\begin{itemize}
\item[i.] For any  $p$ and $q$ such that $p+q \leq 1$, we have
\begin{align}\label{eq:Asymp1}
C=1-(p+q)\frac{\log{\ell}}{\ell}+O(\ell^{-1})\,;
\end{align}
\item[ii.]  When $q=0$ and $p=O(\ell^{-1})$ we have
\begin{align} \label{eq:AsympLowDel}
{L}^{0.5}=&1+(p/\ell)\log{(p/\ell)}-K_1\cdot (p/\ell)+O(\ell^{-3})
\end{align}
where $K_1 \defeq
\log(2e)-\sum_{j=1}^{\infty}{\frac{j
\log{j}}{2^{j+1}}} \simeq 1.15416377$;
\item[iii.] When $p=0$ and $q=O(\ell^{-1})$ we have
\begin{align}\label{eq:AsympLowDup}
{L}^{0.5}=&1+(q/\ell)\log(q/\ell)+K_2 \cdot (q/\ell)+O(\ell^{-3})
\end{align}
where $K_2 \defeq \sum_{j=1}^{\infty}{\frac{j \log{j}}{2^{j+1}}}-\log(\frac{e}{2}) \simeq 0.84583623$.
\end{itemize}
\end{corollary}
Note that the first three terms on the right-hand side of \eqref{eq:AsympLowDel} correspond to the first terms in the asymptotic expansion of the capacity of the i.i.d. deletion channel with deletion probability~$p/\ell$.

\section{Proofs}\label{section3}
We denote by $p_d$  and $p_i$ the unconditional probabilities of deletion and duplication, respectively, of each bit within a block of length $\ell$, {\it{i.e.}},
$$p_d\defeq p/\ell \quad \;p_i\defeq q/\ell\,.$$ Also, we denote by $n_r(\xb)$ the number of runs in a sequence~$\xb$.
\subsection{Proof of ~Proposition \ref{Thm:CapacityMarkov}}\label{sb:MarkLowbound}
\subsubsection{Lower bound}
The left-hand side of \eqref{eq:MarLowDelDup} holds because of \eqref{eq:dobru}. 
\subsubsection {Upper bound}\label{sb:UP}
For any length $\ell$ output sequence $\yb$, 
we have $P_Y(y^{\ell})=(1-p-q)P_{X}(y^{\ell})$ and
$Q(y^{\ell}|x^{\ell})=(1-p-q)$. For a length
$\ell-1$ (respectively, $\ell+1$) output sequence $\yb$, resulting from a one-bit deletion (respectively, duplication) in
the $i$-th run of $x^{\ell}$, we have
$Q(\yb|x^{\ell})=\frac{p\cdot r_i}{\ell}$ (respectively, $\frac{q\cdot r_i}{\ell}).$ Thus, we can write
\begin{align}\label{mlag}
{I}(&X^{\ell};\Yb(X^{\ell}))=H(\Yb(X^{\ell}))-H(\Yb(X^{\ell})|X^{\ell}) \nonumber \\
&=(1-p-q)H(X^{\ell})\nonumber \\
&+(p+q)\sum_{\xb \in \{0,1\}^{\ell}}{P_X(\xb)\sum_{i\in\{1,...,n_r(\xb)\}}{\frac{r_i}{{\ell}}\cdot \log\frac{r_i}{{\ell}}}}\nonumber \\
&-\sum_{|\yb|={\ell}-1}{P_Y(\yb)\log P_Y(\yb)}-\sum_{\yb: |\yb|=n_r(\yb)}{P_Y(\yb)\log P_Y(\yb)} \nonumber \\
&+p\log{p}+q\log{q} \,,
\end{align}
 The sum of the first two terms on the right-hand side of the second equality is
a concave function  of $P_X$. By the Lagrange multipliers
method one deduces that the maximum is attained for the distribution
\begin{align}
P^*_X(x^{\ell})=\frac{2^{-\frac{p+q}{1-p-q}{\hat{H}(\rb(x^{\ell}))}}}{\sum_{\xb\in\{0,1\}^\ell}{2^{-\frac{p+q}{1-p-q}{\hat{H}(\rb(\xb))}}}}\,. \nonumber
\end{align}
Maximizing separately the third and the fourth terms on the right-hand side of the second equality in
\eqref{mlag} under the constraints $\sum_{|\yb|=\ell-1}P_Y(\yb)=p$ and $\sum_{\yb: |\yb|=n_r(\yb)}P_Y(\yb)=q$
is similar to entropy maximization and the maximums are 
achieved by the distributions $$P^{**}_Y(y^{\ell-1})=\frac{p}{2^{\ell-1}}\quad \text{and}\quad \;P^{***}_Y(y^{\ell+1})=\frac{q}{2^{\ell+1}-2}\,,$$
respectively.

Substituting distributions $P^*_X$, $P^{**}_Y$, and $P^{***}_Y$ on the right-hand side of the second equality in \eqref{mlag} we obtain $U$.

\subsection{Proof of Theorem ~\ref{Thm:MarLowBoundAsymp}}
\begin{itemize}

\item[i.]
This part of the theorem is obtained by deriving the asymptotic behavior of \eqref{eq:LowUnifDelDup} as $\ell \to \infty$. To do this, we need the following lemma: 
\begin{lemma} \label{lem:asymp} 
For any positive $s,t$ such that $s+t=1$, we have:
\begin{align}\label{eq:S_2_n}
\sum_{k=1}^{n}{\binom{n}{k}  s^k t^{n-k} k \log{k}}=&sn \log(sn)+t \log{e}+\frac{s-1}{2}\nonumber \\
&+O(\frac{1}{n})\,.
\end{align}
\end{lemma}
\begin{IEEEproof}[Proof of Lemma~\ref{lem:asymp}]
 This lemma is proved via the moment generating function method of  \cite{FlajoletSing}. For any sequence of real numbers $\{f_k\}$, the Bernoulli transform of $f_k$ is defined as
\begin{align*}
S_n\defeq \sum_{k=0}^{n}{\binom{n}{k} f_k s^k t^{n-k}} \,,
\end{align*}
Further, for $f_k$ and its Bernoulli transform $S_n$, the generating functions are defined by 
\begin{align}
f(z) \defeq \sum_{k \geq 1}{f_k z^k}\quad \text{and}\quad
S(z) \defeq \sum_{n \geq 1}{S_n z^n}\,,\nonumber 
\end{align}
respectively.

It is easy to check (see \cite{FlajoletSing}) that $f$ and $S$ satisfy 
\begin{align*}
S(z)=\frac{1}{1-tz}f(\frac{sz}{1-tz})\,.
\end{align*}
Now we consider two sequences of real numbers $f_k^{(1)} \defeq \log{k}$ and $f_k^{(2)} \defeq k \log{k}$, $k\geq 1$. For $f_k^{(i)}$ and $i \in \{1,2\}$, we denote the Bernoulli transform, generating function, and generating function of the Bernoulli transform by $S_n^{(i)}$, $f^{(i)}(z)$, and $S^{(i)}(z)$, respectively. Also, we denote by $g'$ the first derivative of a function $g$. 

It is easy to check that $f^{(2)}(z)=z\cdot(f^{(1)})'(z)$ which implies that
\begin{align*}
S^{(2)}(z)=-tz S^{(1)}(z)+(1-tz)z (S^{(1)})'(z)\,.
\end{align*}
Now, from \cite[Propostion 1]{FlajoletSing}, we know that
\begin{align*}
S^{(1)}_n=\log{sn}+\frac{s-1}{2sn}+O(\frac{1}{n^2})\,.
\end{align*}
Denote by $[z^n]A(z)$ the $n$-th coefficient of a generating function $A(z)$. Since $[z^n] z^k S(z)=S_{n-k}$ and $[z^n] S'(z)= (n+1)S_{n+1}$, we obtain 
\begin{align*}
S_n^{(2)}&=-t[\log{(s(n-1))}+\frac{s-1}{2s(n-1)}+O(\frac{1}{n^2})]\nonumber \\
&\vspace{0.2cm}+n[\log{(sn)}+\frac{s-1}{2sn}+O(\frac{1}{n^2})]\nonumber\\
&\vspace{0.2cm}-t(n-1)[\log{s(n-1)}+\frac{s-1}{2s(n-1)}+O(\frac{1}{n^2})]\nonumber\\
&=sn \log(sn)+t \log{e}+\frac{s-1}{2}+O(\frac{1}{n})\,.
\end{align*}
Since $S_n^{(2)}$ corresponds to the left-hand side of \eqref{eq:S_2_n} the proof is complete.  
\end{IEEEproof}
For any $p$ and $q$, as $\ell \to \infty$, we have
\begin{align}\label{eq:asym_t1}
&\frac{p+q}{\ell^2}\sum_{j=1}^{\ell-1}{\frac{\ell-j+3}{2^{j+1}}j \log{j}}\nonumber\\
&\hspace{1.7cm}=\frac{(p+q)K}{\ell}+(p+q)O(\ell^{-2})\,,
\end{align}
where $K$ is defined as
\begin{align*}
K \defeq \lim_{\ell \to \infty}\sum_{j=1}^{\ell}{2^{-(j+1)} j \log{j}}\,.
\end{align*}
Also, we have
\begin{align}\label{eq:asym_t2}
&\frac{ q}{\ell^2 \cdot 2^{\ell-1}}\sum_{m=1}^{\ell}{m  \binom{\ell}{m}} \log{m} \nonumber \\
&\hspace{1cm}=\frac{2\cdot q}{\ell^2}\sum_{m=1}^{\ell}{(0.5)^m(0.5)^{\ell-m}\binom{\ell}{m} m\log{m}}\nonumber \\
&\hspace{1cm}\overset{a}{=}\frac{2\cdot q}{\ell^2}[\frac{\ell}{2} \log{\frac{\ell}{2}}]+q\,O(\ell^{-2})\nonumber \\
&\hspace{1cm}=-\frac{q}{\ell}+\frac{q}{\ell} \log{\ell}+q\,O(\ell^{-2})\,,
\end{align}
where $a$ follows from Lemma \ref{lem:asymp} by setting $s=t=0.5$. By substituting  \eqref{eq:asym_t1} and \eqref{eq:asym_t2} into \eqref{eq:LowUnifDelDup} we obtain \eqref{eq:asymtilde}.
\item[ii.]
Since the runlengths of a length $\ell$ sequence are between $1$ and $\ell$, we have $\hat{H}(r(x^{\ell})) \leq \log{\ell}$. If we assume that $(p+q) \log{\ell} \to 0$, we can use Taylor's expansion of $2^{-x}$ around $x=0$ to get
\begin{align}
2^{-\frac{p+q}{1-p-q}\hat{H}(r(x^{\ell}))}=&1-\frac{(p+q)}{(1-p-q)\log{e}}\hat{H}(r(x^{\ell}))\nonumber \\
&+O((p+q)^2 \ell^2)\,.
\end{align}
Thus, we have
\begin{align}
&\sum_{x^\ell}{2^{-\frac{p+q}{1-p-q}\hat{H}(r(x^{\ell}))}}=2^{\ell}(1-O((p+q)^2 \ell^2))\nonumber \\
&-\frac{(p+q)}{(1-p-q)\log{e}}\sum_{x^\ell}{\hat{H}(r(x^{\ell}))}\,.
\end{align}
Now, we establish the asymptotic behavior of $\sum_{x^\ell}{\hat{H}(r(x^{\ell}))}$. Denoting by $n(\ell,j)$, the number of times a run with length of $j$ appears in all length  $\ell$ sequences, we have
\begin{align}
\sum_{x^\ell}{\hat{H}(r(x^{\ell}))}&=-\sum_{j=1}^{\ell}{n(\ell,j)\frac{j}{\ell}\log{(\frac{j}{\ell})}}\nonumber \\
&\overset{a}{=}-\sum_{j=1}^{\ell-1}{2^{\ell-j-1}\frac{\ell-j+3}{\ell} j \log{\frac{j}{\ell}}} \nonumber \\
&=-2^{\ell}\Big( \sum_{j=1}^{\ell-1}{{2^{-j-1}\frac{\ell-j+3}{\ell} j \log{{j}}}}\nonumber \\
&\hspace{0.5cm}-\log{{(\ell)}}\cdot \sum_{j=1}^{\ell-1}{{2^{-j-1}\frac{\ell-j+3}{\ell} j}}\Big)\nonumber \\
&\overset{b}{=}2^{\ell} (\log{\ell}-K)\,,
\end{align}
where $a$ follows from \cite[Proposition 2]{Rahmati11} and where $b$ follows from $\sum_{j=1}^{\infty}{\frac{j}{2^{j+1}}}=1$. Therefore, we have
\begin{align} \label{eq:asymup1}
&\log{(\sum_{x^\ell}{2^{-\frac{p+q}{1-p-q}\hat{H}(r(x^{\ell}))}})}\nonumber \\
&=\ell+\log{(1-\frac{p+q}{(1-p-q)\log{2}}(\log{\ell}-K)}\nonumber \\
&\hspace{0.2cm}+O(((p+q)\log{\ell})^2)) \nonumber \\
&=\ell-\frac{p+q}{1-p-q}(\log{\ell}-K)+O(((p+q)\log{\ell})^2)\,.
\end{align} 
By substituting \eqref{eq:asymup1} in \eqref{eq:Up_analytical}, we  obtain \eqref{eq:asymup}.
\item[iii.] The capacity expansion in \eqref{eq:Asymp_cap} follows from \eqref{eq:asymtilde}, \eqref{eq:asymup}.
\end{itemize}
\subsection{Proof of Corollary~\ref{cor:asymseg}}
\begin{itemize}
\item[i.] Since $\frac{H_b(p,q)}{\ell}=O(\ell^{-1})$, we have
\begin{align*}
L^{0.5}=1-\frac{(p+q)\log{\ell}}{\ell}+O(\ell^{-1})\,.
\end{align*}
Also \eqref{eq:generalUB}, \eqref{eq:MarLowDelDup} , \eqref{eq:asymup} imply that $C$ is upper bounded by $1-\frac{(p+q)\log{\ell}}{\ell}+O(\ell^{-1})$. The proof is complete.
\item[ii.] We expand $(1-p)\log{(1-p)}$ around $p=0$ to obtain 
\begin{align*}
L^{0.5}=&1-\frac{p}{\ell}\log{\ell}-\frac{p}{\ell}+\frac{p}{\ell} \log{(\frac{p}{\ell}\ell)}+(\frac{p}{\ell})K\nonumber \\
&\vspace{0.2cm}+\frac{(1-p)}{\ell}(-p+O(p)^2)\log{e}+p O(\ell^{-2})\nonumber \\
&=1+p_d \log{p_d}-(\log{2e}-K)p_d+O(\ell^{-3})\,.
\end{align*}
\item[iii.] The proof is similar to the previous case. 
\end{itemize}

\appendix
For any $p,q,\alpha\in [0,1]$ such that $p+q\leq 1$,
and any integer $\ell>1$
we have
\begin{align}\label{eq:L_Markov}
{L}^{\alpha}_{SI} & =\frac{1+(1-p)(\ell-1)H_b(\alpha)+(p+q)(1-\alpha)^{\ell-1} \log{\ell}}{\ell} \nonumber\\
&-\frac{p}{\ell^2(1-\alpha)}\Big[\Big(2\alpha^3(\ell-2)-(\ell^2+\ell-6)\alpha^2\nonumber \\
&\hspace{1cm}+(\ell^2-3\ell-2)\alpha+2\ell \Big)\log{\alpha} \nonumber \\
&\hspace{1cm}+\Big( -2\alpha^3(\ell-2)+\alpha^2(\ell^2+\ell-6)\nonumber \\
&\hspace{1cm}-2\alpha(\ell^2-2\ell-1)+\ell(\ell-3)\Big)\log(1-\alpha)\Big]\nonumber\\
&-\frac{p \alpha^2 (1-\alpha)^{\ell-3}}{\ell^2} \sum_{m=0}^{\ell-2}\Big[\binom{\ell-2}{m}(\beta+\gamma+\gamma m) \times \nonumber \\
&\hspace{2cm}(\frac{\alpha}{1-\alpha})^m \log{(\beta+\gamma+\gamma m)}\Big] \nonumber \\
&-\frac{q \alpha^{\ell}}{\ell^2 (1-\alpha)}\sum_{m=1}^{\ell}{\binom{\ell}{m}(\frac{\alpha}{1-\alpha})^{-m}m\log{m}}\nonumber \\
&+\frac{p+q}{\ell^2}\sum_{m=2}^{\ell}{m \alpha^ {m-1} (1-\alpha)^{\ell-m}}\times\nonumber \\
&\hspace{2cm}\Big(\sum_{k=1}^{\ell-m+1}{\binom{\ell-k-1}{m-2} k\log{k}}\Big)\,,
\end{align}
 where 
\begin{align*}
\gamma\defeq \frac{1-2\alpha}{\alpha^2}\quad
\text{and}\quad
\beta\defeq\frac{\ell-1+(\alpha^2-\alpha)(2\ell-4)}{\alpha^2}\,.
\end{align*}
When $\alpha=1/2$ the above expression reduces to
\begin{align}\label{eq:LowUnifDelDup}
{L}_{SI}^{0.5}=& 1-\frac{p}{\ell}-\frac{ q}{\ell^2 \cdot 2^{\ell-1}}\sum_{m=1}^{\ell}{m  \binom{\ell}{m}} \log{m}\nonumber\\ 
&-(p-\frac{p+q}{2^{\ell-1}})\frac{\log{\ell}}{\ell}\nonumber \\
&+\frac{p+q}{\ell^2 }\sum_{j=1}^{{\ell-1}}{{\frac{({\ell}-j+3)}{2^{j+1}}\times j \log{j}}} \,.
\end{align}

\begin{IEEEproof}
In order to prove \eqref{eq:L_Markov}, we need the following lemmas. 
\begin{lemma} \label{lem:series} 
For any integer $n \geq 1$, we have
\begin{align}\label{eq:Series}
\sum_{k=0}^{n}{\binom{n}{k}k \cdot t^k}&=n(1+t)^{n-1}t\nonumber \\
\sum_{k=0}^{n}{\binom{n}{k} k^2 \cdot t^k}&=n(1+t)^{n-1}t+n(n-1)(1+t)^{n-2}t^2\,.
\end{align}
\end{lemma}
\begin{IEEEproof}[Proof of Lemma~\ref{lem:series}]
The first and second equations can be obtained by taking the first and second derivatives  with respect to $t$ of the  Binomial equation
$$\sum_{k=0}^{n}{\binom{n}{k}t^k}=(1+t)^{n}\,.$$
\end{IEEEproof}
\begin{lemma} \label{lem:numkml} 
\begin{itemize}
\item The number of length $\ell$ sequences containing $m$ runs is
\begin{align}\label{eq:numml}
n'(\ell,m)=2\binom{\ell-1}{m-1}\,.
\end{align}
\item The number of length $k$ runs among all length $\ell$ sequences containing $m$ runs is
\begin{align}\label{eq:numkml}
 n''(k,m,\ell) = \left\{ 
  \begin{array}{l l}
  2 &  \text{if $m=1,k=\ell$}\\
    2m\binom{\ell-k-1}{m-2} & \text{if $m\geq2, k \leq \ell-m+1$}\\
    0 &  \text{otherwise}\\
  \end{array} \right.
 \end{align}
\end{itemize}

\end{lemma}
\begin{IEEEproof}[Proof of Lemma~\ref{lem:numkml}]
\begin{itemize}
\item The number of length $\ell$ sequences containing $m$ runs is twice the number of positive integer solutions of equation
 \begin{align}\label{eq:lineq}
r_1+\cdots+r_m=\ell\
\end{align}
which is  \cite{Mahmoudvand}
$$\binom{\ell-1}{m-1}\,.$$

\item Since the only two sequences containing $1$ run are the all-zero and all-one sequences we have $n''(\ell,1,\ell)=2$. The number of runs of length $k$ among all length $\ell$ sequences containing $m$ runs is twice the number of times $k$ appears in the solution set of \eqref{eq:lineq}.
The number of times that the first run has length $k$ is twice the number of positive integer solutions of $r_2+\cdots+r_m=\ell-k$. Therefore, the number of times a run of length $k$ appears in all length $\ell$ sequences containing $m$ runs is equal to $2m\binom{\ell-k-1}{m-2}$.  
\end{itemize}
\end{IEEEproof}
We write $L_{SI}^{\alpha}$ as  $$L_{SI}^{\alpha}=\frac{H(\tilde{\Yb})-H(\tilde{\Yb}|X^{\ell}(\alpha))}{\ell}\,.$$   
First, we calculate $H(\tilde{\Yb}(X^{\ell}(\alpha)))$. To compute this entropy, we need to calculate the probabilities of all output sequences. We classify the output sequences according to their lengths. For length $\ell$ sequences, we have $$P_Y(y^\ell)=(1-p-q)P_X(y^\ell)\,,$$ which results in
\begin{align}\label{eq:Hyell}
-\sum_{y^\ell}{P_Y(y^\ell)\log{P_Y(y^\ell)}}&=(1-p-q)H(X{^\ell}(\alpha))\nonumber\\
&\hspace{0.75cm}-(1-p-q)\log(1-p-q)
\end{align}
where the input block entropy is given by
\begin{align}\label{eq:EntXMar}
H(X^{\ell}(\alpha))=1+(\ell-1)H_b(\alpha)\,.
\end{align}
Now, we turn to output sequences of length $\ell-1$. For any $\alpha \in [0,1]$ and integers $\ell\geq 1$ and $m \leq \ell$, we define
\begin{align}\label{eq:defoff}
f(\ell,m,\alpha) \defeq 0.5 \, (1-\alpha)^{\ell-m-1} {\alpha}^m\,.
\end{align}
The probability of any sequence generated by a first-order Markov process is a function of the number of its transitions.\footnote{Number of transitions of a sequence is the number of times its two consecutive bits differ} Since the number of transitions of a sequence $\xb$ is equal to $n_r(\xb)-1$, for any length $\ell$ sequence generated by \eqref{eq:DisMarkov}, we can write
\begin{align}
P_X(\xb)=f(\ell,n_r(\xb)-1,\alpha)\,.
\end{align} 
To calculate $P_{Y}(y^{\ell-1})$, we need to calculate the probability of each of its length $\ell$ super-sequences.\footnote{$\xb$ is super-sequence of $\yb$ if $\yb$ is a subsequence of $\xb$} A length $\ell$ super-sequence of $\yb$ can be generated by inserting one bit into $y^{\ell-1}$ in one of the following ways: 
\begin{itemize}
\item  Insert one zero (one) to one of its runs of zeros (ones). The number of distinct super-sequences generated under this scenario is equal to $n_r(\yb)$. Let $\xb'$ be sequence $\yb$ with one bit inserted in its $i$-th run. Hence we have $Q(\yb|\xb')=(r_i(\yb)+1) \cdot p_d$. Also, note that for any such $\xb'$ we have $n_r(\yb)=n_r(\xb')$ and thus, $P_X(\xb')=f(\ell,n_r(\yb)-1,\alpha)$.
\item Insert one opposite bit at one of its ends. The number of possible super-sequences generated under this scenario is $2$. For any such super-sequences $\xb''$ we have $Q(\yb|\xb'')=p_d$. Also, note that $n_r(\xb'')=n_r(\yb)+1$ and thus $P_X(\xb'')=f(\ell,n_r(\yb),\alpha)$. 
\item Insert one opposite bit inside of one of its runs.  Since for any sequence of length $\ell-1$, there are $\ell+1$ super-sequences of length $\ell$, the number of possible super-sequences generated under this scenario is $\ell-n_r(\yb)-1$. For any such $\xb'''$ we have $Q(\yb|\xb''')=p_d$ and $P_X(\xb''')=f(\ell,n_r(\yb)+1,\alpha)$. 
\end{itemize}
Therefore, for any $\yb \in \{0,1\}^{\ell-1}$, we have
\begin{align}\label{eq:Pydel1}
P_Y(\yb)&=\sum_{x^{\ell}}{P_X(x^\ell)Q(\yb |x^\ell)}  \nonumber \\
&=\Big(\beta+\gamma\cdot n_r(\yb)\Big) f\Big(\ell,n_r(\yb)+1,\alpha \Big)p_d\,.
\end{align}
Hence, we have
\begin{align}\label{eq:Pydel2} 
&-\sum_{\yb \in \{0,1\}^{\ell-1}}{P_Y(\yb) \log{P_Y(\yb)}} \nonumber \\
&=-(p_d \log{p_d})\sum_{\yb}{\Big(\beta+\gamma \cdot n_r(\yb)\Big) f\Big(\ell,n_r(\yb)+1,\alpha \Big)} \nonumber \\
&\hspace{0.35cm}-p_d \sum_{\yb}{\Big(\beta+\gamma \cdot n_r(\yb)\Big) f\Big(\ell,n_r(\yb)+1,\alpha \Big)} \times \nonumber   \\ &\hspace{1.8cm}{\log{\Big[ \Big(\beta+\gamma \cdot n_r(\yb)\Big) f\Big(\ell,n_r(\yb)+1,\alpha \Big)\Big ]}} \nonumber \\
&=-(p_d \log{p_d})\cdot A_1-p_d \cdot (A_2+A_3) \,,
\end{align}
where
\begin{align}\label{eq:Pydel3}
A_1& \defeq \sum_{\yb}{\Big(\beta+\gamma \cdot n_r(\yb)\Big) f\Big(\ell,n_r(\yb)+1,\alpha \Big)} \nonumber \\
&=\ell \,, 
\end{align}
\begin{align}\label{eq:Pydel5}
A_2& \defeq \sum_{m=1}^{\ell-1}\Big[n'(\ell-1,m)(\beta+\gamma m)f(\ell,m+1,\alpha) \times \nonumber \\
&\hspace{1.5cm}\log{(\beta+\gamma m)} \Big] \nonumber \\
&=\alpha^2 (1-\alpha)^{\ell-3} \sum_{m=0}^{\ell-2}\Big[\binom{\ell-2}{m}(\beta+\gamma+\gamma m) \times \nonumber \\
&\hspace{1.5cm}(\frac{\alpha}{1-\alpha})^m \log{(\beta+\gamma+\gamma m)}\Big]\,,
\end{align}
\begin{align}\label{eq:Pydel6}
A_3& \defeq \sum_{m=1}^{\ell-1} \Big[\binom{\ell-2}{m-1}(\beta+\gamma m)(1-\alpha)^{\ell-m-2}{\alpha}^{m+1} \nonumber \\ &\hspace{1.5cm} \times \log{\Big(0.5 (1-\alpha)^{\ell-m-2}{\alpha}^{m+1}\Big)}\Big] \nonumber \\
&=B_1+B_2-B_3\,,
\end{align}
with
\begin{align}\label{eq:Pydel7}
B_1&\defeq {\alpha (1-\alpha)^{\ell-2} \log{\alpha} } \times \nonumber \\
&\hspace{1cm} \sum_{m=1}^{\ell-1}\Big[\binom{\ell-2}{m-1} (\beta+\gamma m)\times (\frac{\alpha}{1-\alpha})^m (m+1)\Big] \nonumber \\
&=\frac{\log{\alpha}}{1-\alpha}[2\alpha^3(\ell-2)-(\ell^2+\ell-6)\alpha^2\nonumber \\
&\hspace{1cm}+(\ell^2-3\ell-2)\alpha+2\ell]\,, 
\end{align}
\begin{align}\label{eq:Pydel8}
B_2 \defeq &\alpha (1-\alpha)^{\ell-2} \log({1-\alpha})\times \nonumber \\ 
&\sum_{m=1}^{\ell-1}\Big[\binom{\ell-2}{m-1} (\beta+\gamma m)(\frac{\alpha}{1-\alpha})^m (\ell-m-2) \Big] \nonumber \\
&=\frac{\log(1-\alpha)}{1-\alpha}[-2\alpha^3(\ell-2)+\alpha^2(\ell^2+\ell-6)\nonumber\\
&\hspace{0.75cm}-2\alpha(\ell^2-2\ell-1)+\ell(\ell-3)]\,,
\end{align}
\begin{align}\label{eq:Pydel9}
B_3&\defeq\sum_{m=1}^{\ell-1} \Big[\binom{\ell-2}{m-1}(\beta+\gamma m)(1-\alpha)^{\ell-m-2}{\alpha}^{m+1}\Big] \nonumber \\
&=\ell \,.
\end{align}
Now, we consider the length $\ell+1$ output sequences. Obviously, for the alternating sequences ({\it{i.e.}}, $\yb$ such that $|\yb|=n_r(\yb)$) of length $\ell+1$, we have $P_Y(\yb)=0$. Denoting by $\mathcal{Y}^*$ the set of length $\ell+1$ non-alternating sequences, for any $\yb \in \mathcal{Y}^*$ the duplicated bit can be found in one of the runs of $\yb$ with a length greater than $1$. Hence, for any $\yb \in \mathcal{Y}^*$, we have 
\begin{align*}
P_Y(\yb)&=\sum_{j:r_j(\yb)>1}{(r_j(\yb)-1)p_i\cdot f(\ell,n_r(\yb)-1,\alpha)}\nonumber \\
&=(\ell+1-n_r(y))p_i\cdot f(\ell,n_r(\yb)-1,\alpha)\,,
\end{align*}
where the second equality follows from the fact that duplication error can not create a new run in the received sequence. 
Thus, we have
\begin{align}\label{eq:Pyins2} 
&-\sum_{\yb \in \mathcal{Y}^*}{P_Y(\yb)\log{P_Y(\yb)}}=q(1-\log{q})+q(\ell-1)H_b(\alpha)\nonumber\\
&\hspace{0.5cm}+q\log{\ell}-\frac{q}{\ell}
{\frac{\alpha^{\ell}}{1-\alpha}}\sum_{m=1}^{\ell}{\binom{\ell}{m}(\frac{\alpha}{1-\alpha})^{-m}m\log{m}}\,.
\end{align}

Now, we turn to $H(\Yb(X^\ell)|X^{\ell})$. We have
\begin{align}\label{eq:HYX1}
H&(\Yb(X^\ell)|X^\ell)=H_b(p,q)+(p+q)\log{\ell} \nonumber \\&-\frac{p+q}{\ell}\sum_{\xb \in \{0,1\}^\ell}{P_X(\xb)\sum_{i=1}^{n_r(\xb)}{r_i(\xb) \log{r_i(\xb)}}}\,.
\end{align}
Denoting by $n''(k,m,\ell)$ the number of times a run of length $k$ appears in all possible length $\ell$ sequences containing $m$ runs we have
\begin{align}\label{eq:HYX2}
&\sum_{\xb \in \{0,1\}^\ell}{P_X(\xb)\sum_{i=1}^{n_r(\xb)}{r_i(\xb) \log{r_i(\xb)}}}\nonumber \\
&=\sum_{m=1}^{\ell}{\sum_{\xb: n_r(\xb)=m}{P_X(\xb)\sum_{i=1}^{n_r(\xb)}{r_i(\xb) \log{r_i(\xb)}}}}\nonumber \\
&=\sum_{m=1}^{\ell}{f(\ell,\alpha,m-1)\sum_{\xb: n_r(\xb)=m}{\sum_{i=1}^{m}{r_i(\xb) \log{r_i(\xb)}}}} \nonumber \\
&=\sum_{m=1}^{\ell}{f(\ell,\alpha,m-1)\sum_{k=1}^{\ell-m+1}{n''(k,m,\ell)\cdot k\log{k}}} \nonumber \\
&=\sum_{m=2}^{\ell}{m {} f(\ell,\alpha,m-1)\sum_{k=1}^{\ell-m+1}{2 \binom{\ell-k-1}{m-2} \cdot k\log{k}}} \nonumber \\
&\hspace{1cm}+(1-\alpha)^{\ell-1}\ell \log{\ell}
\end{align}
where the last equality follows from Lemma~\ref{lem:numkml}. Putting \eqref{eq:Pydel1}, \eqref{eq:Pydel2}, \eqref{eq:Pydel3}, \eqref{eq:Pydel5}, \eqref{eq:Pydel6}, \eqref{eq:Pydel7}, \eqref{eq:Pydel8}, \eqref{eq:Pydel9}, \eqref{eq:Pyins2}, \eqref{eq:HYX1}, \eqref{eq:HYX2} together, we get  ${L}_{SI}^{\alpha}$.

\end{IEEEproof}
\end{document}